\documentclass{article}
\usepackage{arxiv}

\usepackage[utf8]{inputenc} % allow utf-8 input
\usepackage[T1]{fontenc}    % use 8-bit T1 fonts

\usepackage{bm}
\usepackage{mathtools}
\usepackage{booktabs}
\usepackage{subcaption}
\usepackage{siunitx}
\usepackage{amssymb,amsmath,amsfonts,bigints}
\usepackage{hyperref}
\usepackage[capitalise]{cleveref}

\usepackage{algorithm}
\usepackage{algpseudocode}
\usepackage{xcolor}

\usepackage{tikz}
\usetikzlibrary{calc,arrows.meta,positioning}

\usepackage{glossaries}
\glsdisablehyper

\title{Sample-Free Safety Assessment of Neural Network Controllers via Taylor Methods}

\author{ 
    \href{https://orcid.org/0000-0003-4137-5688}{\includegraphics[scale=0.06]{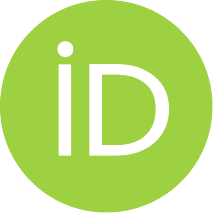}\hspace{1mm}Adam Evans}\thanks{Postdoctoral Fellow, Te Pūnaha Ātea – Space Institute} 
    \hspace{1mm} and \hspace{0.3mm} 
    \href{https://orcid.org/0000-0002-3516-6428}{\includegraphics[scale=0.06]{orcid.pdf}\hspace{1mm}Roberto Armellin}\thanks{Professor, Te Pūnaha Ātea – Space Institute} \\
	University of Auckland\\
	Auckland 1010\\
    New Zealand\\
}

\date{}

\renewcommand{\headeright}{}
\renewcommand{\undertitle}{}
\renewcommand{\shorttitle}{SAMPLE-FREE SAFETY ASSESSMENT OF NEURAL NETWORK
CONTROLLERS}

\hypersetup{
pdftitle={Sample-Free Safety Assessment of Neural Network Controllers via Taylor Methods},
pdfsubject={q-bio.NC, q-bio.QM},
pdfauthor={Adam Evans, Roberto Armellin},
}

\begin{document}
\maketitle

\newacronym{OCP}{OCP}{optimal control problem}
\newacronym{DA}{DA}{differential algebra}
\newacronym{ADS}{ADS}{automatic domain splitting}
\newacronym{DNN}{DNN}{deep neural network}
\newacronym{ODE}{ODE}{ordinary differential equation}
\newacronym{SIREN}{SIREN}{sinusoidal representation network}
\newacronym{SOI}{SOI}{sphere of influence}
\newacronym{TPBVP}{TPBVP}{two-point boundary value problem}
% \vspace{-1em}

\begin{abstract}
In recent years, artificial neural networks have been increasingly studied as feedback
controllers for guidance problems. While effective in complex scenarios, they lack the
verification guarantees found in classical guidance policies. Their black-box nature
creates significant concerns regarding trustworthiness, limiting their adoption in
safety-critical spaceflight applications. This work addresses this gap by developing a
method to assess the safety of a trained neural network feedback controller via automatic domain splitting and polynomial bounding. The methodology involves embedding the trained neural network into the system’s dynamical equations, rendering the closed-loop system autonomous. The system flow is then approximated by high-order Taylor polynomials, which are subsequently manipulated to construct polynomial maps that project state uncertainties onto an event manifold. Automatic domain splitting ensures the polynomials are accurate over their relevant subdomains, whilst also allowing an extensive state–space to be analysed efficiently. Utilising polynomial bounding techniques, the resulting event values may be rigorously constrained and analysed within individual subdomains, thereby establishing bounds on the range of possible closed-loop outcomes from using such neural network controllers and supporting safety assessment and informed operational decision-making in real-world missions.
\end{abstract}

\keywords{Safety assessment \and Neural network controllers \and Taylor polynomials \and Event map \and Interval arithmetic}

\section{Introduction}
Machine learning capabilities have expanded significantly in recent years, enabling the adoption of \glspl{DNN} as controllers for complex spacecraft maneuvers~\cite{NN_control_orbit_transfer_multiscale,NN_control_asteroid_landing,MyDNNPaper,ESA_NN_Guidance,NN_control_interplanetary_transfers}. Although highly effective, \glspl{DNN} are widely regarded as black-boxes within the scientific community \cite{xAI_WhyShouldITrustYou}, whose internal inference processes are not interpretable by humans---nor, for that matter, the neural networks themselves. Consequently, Explainable Artificial Intelligence has become an established field of research, aiming to explain and interpret predictions of such complex machine learning models \cite{xAI_Brief_Survey,xAI_Brief_Overview}.

For many scientific areas, including astrodynamics, a standard methodology for testing both the performance and behaviour of trained regression \glspl{DNN} is Monte Carlo simulations~\cite{ESA_verification}. However, this methodology has serious limitations. For typical problems with a high-dimensional input space, extensive simulations and significant computational power are required to cover a wide range of possible inputs, but these simulations often miss edge cases. This is further exacerbated by the fact that, to fully certify a controller's safety, a large state space must be explored to assess behaviour across the full operational envelope. More crucially, as Monte Carlo simulations involve random sampling to test a finite set of points, only a subset of the state--space is tested. Thus, even in an extensive Monte Carlo simulation, input points between the tested set are not assessed. This is of critical importance when considering the recent discovery that \glspl{DNN} are vulnerable to \textit{adversarial examples}. First discovered by applying certain imperceptible perturbations to an image, which caused state-of-the-art \glspl{DNN} to misclassify them \cite{AdversExamples}, these are inputs that cause the network to fail in unexpected ways. This vulnerability is of major concern for applying \glspl{DNN} in safety-critical environments \cite{AdversExamples_AttacksAndDefenses}. Consequently, Monte Carlo simulations are unsuitable for safety verification.

The behaviour of \glspl{DNN} can be assessed in a number of different ways. Kernel methods \cite{KernelMethods} map input data into a higher-dimensional space, where a kernel operator quantifies the similarity between input points in the transformed feature space. Kernel analysis \cite{KernelAnalysis} then examines how the network interpolates data in this space, offering insights into the network's decision-making process. However, \glspl{DNN} are approximated by kernel-based methods in the limit of infinite network width, making the network's behaviour effectively linear and thus not fully capturing the inherent nonlinearity. Alternatively, heatmaps are widely used tools that highlight influential areas within the input data, illustrating how the inputs contribute to the network's output. Moreover, heatmaps can be computed in various ways, each providing different interpretations \cite{DNN_Heatmaps}. For example, sensitivity analysis \cite{DNN_sensitivity_analysis,DNN_sensitivity_analysis_book} uses partial derivatives to measure how small changes in the inputs affect the outputs, providing a local explanation for why a certain prediction was made. In contrast, deconvolution methods for heatmap creation \cite{DNN_deconvolution_method} use backpropagation to project the activations back to the input space, highlighting parts of the input that most influenced the prediction. More recently, a method known as \textit{deep Taylor decomposition} was developed \cite{deep_Taylor_decomposition}, where each neuron is treated as a function and expanded as a Taylor polynomial around the current input. A relevance score is assigned to the outputs, serving as a metric for interpretability, and backpropagated through Taylor-approximated neurons to assign relevance to the inputs, ultimately creating a heatmap. Although heatmaps were initially designed for, and subsequently found more practical use in, classification networks, they can also be applied to regression networks. In the context of \gls{DNN} guidance controllers, both kernel and heatmap methods can provide insights into why a particular control policy was predicted. Whilst this information is valuable, the interest lies not in individual predictions, but in the composition of many predictions and the evolution of decision-making over a trajectory. Furthermore, whilst the sensitivity of outputs to inputs is important, ideally one would be able to bound the possible outcomes of using the \gls{DNN} to control the spacecraft, thereby offering formal safety guarantees. For these reasons, the aforementioned methods are inadequate.

Over the past decade, \gls{DA} has emerged as a powerful tool for the prediction, analysis, and control of physical systems. \gls{DA} provides the necessary tools to automatically compute high-order function derivatives and thus Taylor expansions within a computer environment, by substituting the classical implementation of real algebra with a new commutative algebra of Taylor polynomials. Initially developed to accurately describe beam dynamics within particle accelerators~\cite{Berz_DA_First_Description}, \gls{DA} subsequently found great application in astrodynamics to describe the flow of spacecraft through space, mapping initial states to those at a later time~\cite{DAMapping_Conference,DAMapping}. Consequently, this enables one to study the behaviour and stability of motion in the vicinity of a reference solution. With the methods of polynomial inversion and composition available within the \gls{DA} toolbox, it was not long before such polynomial representations of the system flow were extended and manipulated to construct new types of polynomial maps that related initial states and/or control inputs to later points on various event manifolds. High-order Taylor polynomial representations of some event are created, which are composed with the Taylor polynomials of the system flow and further manipulated to create such descriptive maps. As such, they may also be referred to as event maps. The seminal work on this was Ref.~\cite{DA_close_encounter_characterization}, which mapped varying uncertain initial states of an asteroid to the distance of closest approach to a planetary body via a polynomial map. Subsequent work includes the solution to the \gls{TPBVP} defining an \gls{OCP}, which provides the optimal control law to guide a spacecraft to a target for displaced initial states~\cite{DA_TPBVP1,DA_TPBVP2}, and conjunction analysis to predict the probability of collision of two space objects, based on uncertainty in their states~\cite{DA_Conjunction_Analysis}. Thus, \gls{DA} is an ideal tool for systematic safety analysis. Further work in the field of \gls{DA} introduced the technique of \gls{ADS}~\cite{ADS_original}, a method which leverages multiple polynomial sets to accurately handle large uncertainty sets, by automatically and adaptively analysing the truncation error of the Taylor expansions and splitting the domain of interest when required. In the authors previous works, this has been utilised effectively to produce extensive polynomial maps between the error in the initial state of a spacecraft to the optimal control policy to achieve rendezvous with a target~\cite{MyTimeOptPaper,MyFuelOptPaper}. However, the techniques involved in \gls{ADS} can be further leveraged to assess the safety of neural network--based controllers and to support confidence in their real-world deployment.

This work presents a method for assessing the safety of a \gls{DNN} controller within an extensive operating environment via the use of Taylor polynomials and \gls{ADS}. A trained \gls{DNN} controller predicts the optimal control policy based on an input of state. When the \gls{DNN} is embedded into the dynamical system's \glspl{ODE}, the system becomes autonomous. Trajectories generated by the \gls{DNN} controller can then be readily expanded to obtain the high-order Taylor approximation of the system flow. This flow describes possible trajectories based on variations in the input space. By defining an event of interest with a corresponding mathematical description, these polynomials can be manipulated to construct an event map, which maps a state domain into corresponding event values. \gls{ADS} is used to adaptively split the desired domain of interest into a series of subdomains, each containing one of the aforementioned maps. Crucially, these subdomains are created such that each of the individual event maps are accurate over their respective subdomains. Using polynomial bounding techniques, the resulting event values can be constrained and analysed within individual domains. Truncation errors of the event maps are then calculated using similar techniques embedded into the \gls{ADS} framework. Accounting for these errors is critical to ensure the resulting bounds are exhaustive. Consequently, by combining both the event map bounds and truncation errors, safe bounds on the potential outcomes of using the \gls{DNN} controller can be established.

This manuscript is structured as follows. \cref{s:TaylorMethods} outlines the various Taylor methods that shall be utilised in this work. In \cref{s:PolynomialBounding}, the methodology of interval arithmetic and bounding of Taylor polynomials is discussed. The proposed safety verification method is then assessed in two scenarios. In \cref{s:ClohessyWiltshire}, a trained \gls{DNN} controller is evaluated in a planar Clohessy--Wiltshire scenario, demonstrating the verification framework. Following this, in \cref{s:Earth_Mars}, another trained \gls{DNN} controller is assessed during an Earth--Mars transfer. Concluding remarks are provided at the end.

%###############################################################################
%###############################################################################

\section{Taylor Methods}\label{s:TaylorMethods}
\subsection{High-order expansion of the flow}\label{ss:ExpansionOfFlow}
Fundamental to this work is the high-order Taylor expansion of the flow of a system, which defines how points in the state--space `flow' over time due to the \glspl{ODE} of the dynamical system. Suppose the motion of a spacecraft is defined by the following dynamical \gls{ODE}:

\begin{equation}\label{eqn:SystemDynamics}
    \dot{\bm{x}}(t) = \bm{f}(\bm{x}(t),\bm{u}(t),t)
\end{equation}
where $\bm{x} \in \mathbb{R}^{d}$ is the state, $\bm{u} \in \mathbb{R}^{p}$ the control, $t \in \mathbb{R}_+$ the time, and $\bm{f} : \mathbb{R}^{d} \times\mathbb{R}^{p} \times \mathbb{R}_+ \rightarrow \mathbb{R}^{d}$ is a vector field. Now consider a trained neural network controller $\mathcal{N}_{\bm{u}} : \mathbb{R}^{d} \rightarrow \mathbb{R}^{p}$ that takes as input the current spacecraft state vector and returns the control policy vector. Embedding such a control into \cref{eqn:SystemDynamics} (i.e., $\bm{f}(\bm{x}(t),\mathcal{N}_{\bm{u}}(\bm{x}(t)),t)$) renders the system autonomous\footnote{Notably, \textit{any} smooth dynamical system that uses a smooth feedback controller is expandable in terms of Taylor polynomials, and thus can be studied with the methodology proposed in this work. However, many feedback controllers take forms that are readily analysable. Therefore, this work focuses on \glspl{DNN} due to their black-box properties.}; given any initial condition $\bm{x}_0 \in \mathbb{R}^{d}$ at $t_0 \in \mathbb{R}_+$, there exists a unique solution $\bm{x}(t)$ that passes through $\bm{x}(t_0) = \bm{x}_0$. This is the solution to the initial value problem, which can then be used to determine the solution value at any later time. One may then define a function $\bm{\varphi}_{t_0,t_f} : \mathbb{R}^{d} \rightarrow \mathbb{R}^{d}$ that, for any input $\bm{x}_0$, determines the solution value at some later time $t_f$:

\begin{equation}\label{eqn:TransferMap}
    \bm{x}_f = \bm{\varphi}_{t_0,t_f}(\bm{x}_0)
\end{equation}

It should be noted that the final time $t_f$ can also be a variable if desired. As the flow provides a mathematical description of how final states relate to initial states, such a quantity is also widely referred to as a transfer map, or more simply as a map~\cite{BerzModernMM}. This quantity can be readily approximated by high-order Taylor polynomials through \gls{DA} techniques. This is achieved by initialising the initial state within the \gls{DA} framework, and then substituting the algebraic operations to evaluate the ODE within any explicit integration scheme by the corresponding operations in the \gls{DA} framework~\cite{BerzDAinHandbook}. The initialisation of a given initial state is denoted as

\begin{equation}\label{eqn:InitialStateDA}
    \bm{x}_0 \gets \overline{\bm{x}}_0 + \delta \bm{x}_0
\end{equation}
where $\overline{\bm{x}}_0$ corresponds to the original floating-point number for the initial state, and $\delta$ denotes the \gls{DA} variable which can be viewed as a linear perturbation in the initial state. Essentially, these \gls{DA} variables allow the representation of variations within a computer environment without assigning a specific value, enabling the study of system behaviour under arbitrary variations. It is important to emphasise that $\bm{x}_0$ is now a \gls{DA} object and a polynomial---in this case a first-order expansion of the initial state with respect to $\delta \bm{x}_0$. The overline denotes the constant zero-order term, corresponding to the original quantity that serves as the expansion point. The notation in \cref{eqn:InitialStateDA} is adopted to reduce notational clutter and will be used throughout this work.

Using an appropriate integration scheme, for which an efficient, \gls{DA}-compatible version of a 7/8 Dormand–Prince Runge–Kutta method is utilised, provides the following Taylor expansion by integrating $\bm{x}_0$ from $t_0$ to $t_f$ using the dynamics of \cref{eqn:SystemDynamics}:

\begin{equation}\label{eqn:TaylorMap}
    \bm{x}_f = \mathcal{T}_{\bm{x}_f}(\bm{x}_0)
\end{equation}
where $\mathcal{T}_{\bm{x}_f}$ denotes the high-order Taylor expansion of the final state with respect to the initial conditions of \cref{eqn:InitialStateDA}. Comparing \cref{eqn:TaylorMap} with \cref{eqn:TransferMap}, one realises that $\mathcal{T}_{\bm{x}_f}$ is the Taylor approximation of the transfer map, and is thus the desired high-order expansion of the flow.

\subsection{Event Map}\label{ss:EventMap}
Whilst the Taylor polynomial representation of the flow may be analysed to investigate the resulting state dispersions, at varying times of flight, due to the action of the neural network controller, it is not the ideal quantity of interest. Indeed, in some cases it may be illogical to analyse the flow at a fixed time of flight, such as in time-optimal scenarios where flight time varies between trajectories. Furthermore, a standard approach within the literature when training \glspl{DNN} to replicate the optimal control direction, and potentially the throttle magnitude, does not involve learning the remaining time of flight. Instead, the \glspl{ODE} are typically propagated until either a predefined rendezvous tolerance is achieved or a maximum allowed time of flight is reached, whichever comes first. If the spacecraft achieves the rendezvous tolerance, the rendezvous is considered complete, and the thrusters are either switched off or control is transferred to an alternative controller for the next stage. To reliably analyse the performance of such \gls{DNN} controllers, an alternative quantity must be proposed. 

This work utilises \gls{DA} techniques to map initial states to an \textit{event}, which can subsequently be bounded and analysed. An event is compatible if there exists a corresponding mathematical description. Examples of such events include: the distance of closest approach, defined by the derivative of the distance equal to zero; a desired distance from target, defined by the norm of the difference in position vectors equal to the desired value; the time at which a surface is crossed, for instance a set altitude above an irregular shaped asteroid surface, defined as the zero point of some function which describes the irregular surface at the set altitude. The event mapping process consists of two stages: the first is the detection of the event with high accuracy, and the second is to construct a \gls{DA}-based mapping that projects initial state deviations onto the event manifold. 

As the event may or may not occur at any time during a trajectory propagation, the mathematical quantity corresponding to the event must be calculated within the numerical integration of the dynamical \glspl{ODE}. Using a variable step Runge--Kutta-7-8 integration scheme, the necessary calculations for the event are performed at each integration step. Assuming the objective is to identify the minimum event value, any newly computed value that is lower than all previously encountered is stored in memory, overwriting the prior value. In this way, the optimal value is identified over the course of the trajectory. However, since checks are performed at discrete time steps, the event is inherently overstepped. Therefore, for accurate detection, the event time and value typically require refinement with a \gls{DA}-based mapping. The following algorithm refines a detected event time $\overline{t}_{\textrm{e}} \in \mathbb{R}_{+}$ and a corresponding state $\overline{\bm{x}}_{\textrm{e}} \in \mathbb{R}^{d}$ at some event $\mathcal{E}(\bm{x}) \in \mathbb{R}$, where the event condition is defined as $\mathcal{E}(\bm{x}_{\textrm{e}})=0$.

The event time is first initialised within the \gls{DA} framework as 

\begin{equation}\label{eqn:Event_Time_Initialisation}
    t_{\textrm{e}} = \overline{t}_{\textrm{e}} + \delta t_{\textrm{e}}
\end{equation}
 
The detected $\overline{\bm{x}}_{\textrm{e}}$ may then be expanded with respect to $t_{\textrm{e}}$ via a fixed-point iteration method known as the Picard iteration~\cite{PicardIteration}.  The iteration has the form

\begin{equation}\label{eqn:PicardIteration}
    \begin{gathered}
        \bm{x}_{k+1} = \overline{\bm{x}}_{\textrm{e}} + \int_{\overline{t}_{\textrm{e}}}^{t_{\textrm{e}}} \dot{\bm{x}}\left(\bm{x}_{k}(\tau),\tau\right) d\tau
        \\
        \bm{x}_{k=0} = \overline{\bm{x}}_{\textrm{e}}
    \end{gathered}
\end{equation}
where $\tau$ is a dummy variable for the time integration. The upper bound of integration is that of \cref{eqn:Event_Time_Initialisation}. The Picard–Lindelöf theorem states that, if the differential equations $\dot{\bm{x}}$ are Lipschitz continuous in $\bm{x}$, then there exists a unique solution to which the sequence of Picard iterations converges~\cite{PicardBanach}. Thus, after exactly $k$ iterations, the exact $k^\textrm{th}$ order Taylor expansion of the solution flow with respect to the final time is obtained~\cite{BerzModernMM}. Performing this iteration the required amount of times to obtain the desired order expansion, the result may be expressed as

\begin{equation}\label{eqn:EventDetection_StateExpanded}
    \bm{x}_{\textrm{e}} = \mathcal{T}_{\bm{x}_{\textrm{e}}}(t_{\textrm{e}})
\end{equation}

From the above, the quantity corresponding to the event is constructed within the \gls{DA} framework, denoted by

\begin{equation}
    \mathcal{E} = \mathcal{T}_{\mathcal{E}}(t_{\textrm{e}})
\end{equation}

% As the event has been overstepped and thus inaccurately detected, the zero-order term of $\mathcal{E} \neq 0$. This polynomial is then inverted to obtain $t_{\textrm{e}} = \mathcal{T}_{t_{\textrm{e}}}(\mathcal{E})$, after which the event condition $\mathcal{E}=0$ is enforced. The result is the mapped event time, denoted as $t_{\textrm{e}}^*$ which provides the accurate time at which the trajectory intersects the event manifold. The state at the event can then be obtained by evaluating \cref{eqn:EventDetection_StateExpanded} with the mapped event time, hence

As the event has been overstepped, the zero-order term of $\mathcal{E} \neq 0$. This polynomial is then inverted to obtain $t_{\textrm{e}} = \mathcal{T}_{t_{\textrm{e}}}(\mathcal{E})$, after which the event condition $\mathcal{E}=0$ is enforced. The result is the mapped event time, denoted as 

% \begin{equation}\label{eqn:Event_State_and_Time_Refined}
%     \bm{x}_{\textrm{e}} = \mathcal{T}_{\bm{x}_\textrm{e}}\big|_{t_{\textrm{e}}=t_{\textrm{e}}^*} \quad \textrm{where} \quad t_{\textrm{e}}^* = \mathcal{T}_{t_\textrm{e}} \big|_{\mathcal{E}=0}
% \end{equation}

\begin{equation}\label{eqn:Event_State_and_Time_Refined}
    t_{\textrm{e}}^* = \mathcal{T}_{t_\textrm{e}} \big|_{\mathcal{E}=0}
\end{equation}
which provides the accurate time at which the trajectory intersects the event manifold. After the mapping is complete, the inaccurate event time $\overline{t}_{\textrm{e}}$ is overwritten with the accurate time, i.e., $\overline{t}_{\textrm{e}} \gets t_{\textrm{e}}^*$. Additionally, the state at the event can easily be obtained by evaluating \cref{eqn:EventDetection_StateExpanded} with \cref{eqn:Event_State_and_Time_Refined}.

Having acquired the true event time, the flow can now be expanded and all initial state deviations subsequently mapped onto the event manifold. The procedure for this is similar to the above mapping, with the added step of initialising the initial state within the \gls{DA} framework as

\begin{equation}
    \bm{x}_{0} = \overline{\bm{x}}_{0} + \delta \bm{x}_{0}
\end{equation}
in addition to the event time of \cref{eqn:Event_Time_Initialisation}. The initial state is then propagated until $t_{\textrm{e}}^*$ from \cref{eqn:Event_State_and_Time_Refined}, before the Picard iteration is implemented to expand the state at the event with respect to $t_{\textrm{e}}$, thereby providing the flow

\begin{equation}\label{eqn:EventMapping_StateExpanded}
    \bm{x}_{\textrm{e}} = \mathcal{T}_{\bm{x}_\textrm{e}}(\bm{x}_0,t_{\textrm{e}})
\end{equation}

Using these polynomials, the event quantity is then constructed as

\begin{equation}
    \mathcal{E} = \mathcal{T}_{\mathcal{E}}(\bm{x}_0,t_{\textrm{e}})
\end{equation}
where the zero-order term of $\mathcal{E}$ now equals zero due to the previous refinement. Concatenating $\mathcal{E}$ with the initial state variation identity, as in

\begin{equation}\label{eqn:EventMap_FullRankMatrix}
    \begin{bmatrix}
        \mathcal{E} \\ \bm{x}_0   
    \end{bmatrix}
    =
    \begin{bmatrix}
        \mathcal{T}_{\mathcal{E}}(\bm{x}_0,t_{\textrm{e}}) \\ 
        \overline{\bm{x}}_0 + \delta\bm{x}_0
    \end{bmatrix}
\end{equation}
produces a full-rank square matrix, which is subsequently inverted. Extracting the event time polynomial $t_{\textrm{e}} = \mathcal{T}_{t_\textrm{e}}(\bm{x}_0,\mathcal{E})$ and enforcing $\mathcal{E}=0$ then provides the time at which the flow intersects the event manifold. Finally, evaluating \cref{eqn:EventMapping_StateExpanded} with the mapped event time yields the state on the event manifold for any $\bm{x}_0$.

\begin{equation}\label{eqn:Event_State_and_Time_Mapped}
    \bm{x}_{\textrm{e}}^* = \mathcal{T}_{\bm{x}_\textrm{e}}(\bm{x}_0)\big|_{t_{\textrm{e}}=t_{\textrm{e}}^*} \quad \textrm{where} \quad t_{\textrm{e}}^* = \mathcal{T}_{t_\textrm{e}}(\bm{x}_0) \big|_{\mathcal{E}=0}
\end{equation}

\begin{figure}[!t]
    \centering
    \includegraphics[width=0.7\textwidth]{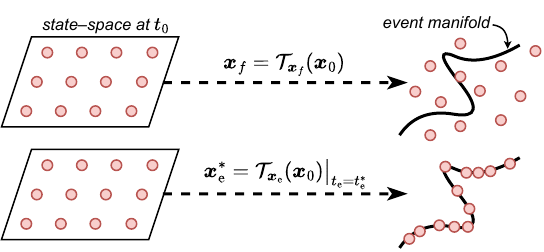}
	\caption{Illustration of the distinction between the transfer map and the event map.}
	\label{fig:transfer_vs_event_map}
\end{figure}

Evaluating the polynomials of \cref{eqn:Event_State_and_Time_Mapped} with any $\bm{x}_0$ returns the state on the event manifold and corresponding time for a trajectory starting at $\bm{x}_0$. The distinction between the transfer map of \cref{eqn:TaylorMap} and the event map of \cref{eqn:Event_State_and_Time_Mapped} is illustrated in \cref{fig:transfer_vs_event_map}. When evaluated with arbitrary initial states, the transfer map provides the corresponding states at some prescribed final time, whereas the event map provides the states on the event manifold. These polynomials may be further manipulated if required---such as computing vector norms or additional quantities---providing corresponding results on the event manifold. Thus, appropriate and relevant quantities are obtained for safety analysis.

\subsection{Automatic Domain Splitting}\label{ss:ADS}
As Taylor polynomials provide strictly local approximations, their validity is constrained by a radius of convergence. This radius generally decreases with increasing nonlinearity of both the dynamics and the description of the event, confining the accurate representation to a small vicinity of the expansion point. Outside this region, the series diverges rapidly, leading to substantial errors that are often exacerbated when higher polynomial orders are used. Consequently, they are unsuitable for investigating mappings further from the expansion point---a capability essential for capturing the full range of initial dispersions and operational uncertainties inherent in astrodynamics settings. In response to this problem, the \gls{ADS} algorithm was developed~\cite{ADS_original}. Whilst an overview is provided here for completeness, the interested reader is referred to the referenced work.

For simplicity, first consider a deterministic scalar function $f: \mathbb{R} \to \mathbb{R}$ that is $\mathcal{C}^{n+1}$ at an expansion point $\overline{x} \in \mathbb{R}$ and its corresponding $n$\textsuperscript{th}-order Taylor approximation $\mathcal{T}_f(x)$. The error in this approximation is given by the remainder term, which is asymptotic in nature and tends to zero faster than any non-zero $n^\textrm{th}$ degree polynomial as $x \to \overline{x}$. Explicit forms of the remainder can be obtained under stronger regularity assumptions on $f$, the most common of which is the Lagrange form that may be proved using the mean value theorem~\cite{Apostol_Calculus}. Crucially, the approximation error is bounded by the highest order term of the $n+1$\textsuperscript{th}-order Taylor polynomial:

\begin{equation}\label{eqn:ADSerror}
    \left| f(x) - \mathcal{T}_f(x) \right| \leq C  \delta x ^{n+1}
\end{equation}
where $\delta x = x - \overline{x}$ and $C \in \mathbb{R}^{*}_{+}$ is the maximum value of $|f^{(n+1)}(\xi)|/(n+1)!$ on the interval $\xi \in [0,\delta x]$. Consider now an uncertainty domain $D_r$ of size $r \in \mathbb{R}^{*}_{+}$ around the expansion point. Given any deviation $\delta x$ that lies inside this domain, the maximum approximation error $e_r \in \mathbb{R}^{*}_{+}$ may be expressed using \cref{eqn:ADSerror} as

\begin{equation}\label{eqn:ADSerror_max}
    \left| f(\delta x) - \mathcal{T}_f(\delta x) \right| \leq C  \delta x ^{n+1} \leq C  r^{n+1} = e_r
\end{equation}
which provides an upper bound on the approximation error within the given domain $D_r$. 

Halving the domain to a new domain $D_{r/2}$ of size $r/2$, the new approximation error bound can be derived using \cref{eqn:ADSerror_max} as 

\begin{equation}\label{eqn:ADSerror_halved}
    \left| f(\delta x) - \mathcal{T}_f(\delta x) \right| \leq C  \delta x ^{n+1} \leq C  \left( \dfrac{r}{2} \right) ^{n+1} = \dfrac{e_r}{2^{n+1}}
\end{equation}

Thus, the maximum approximation error is reduced by a factor of $1/2^{n+1}$. This reduction is amplified as the expansion order $n$ increases. Utilising this, the \gls{ADS} algorithm was developed~\cite{ADS_original} to track the approximation error of the Taylor polynomials and automatically divide domains that possess larger than tolerated approximation errors into two equal subdomains. New Taylor polynomials that are calculated around the centres of each new subdomain exhibit greatly reduced approximation errors within their specified domains due to \cref{eqn:ADSerror_halved}. The problem of when to split a domain is handled by the \gls{ADS} algorithm, which performs the splitting automatically and adaptively, according to when the estimated approximation error of the polynomial exceeds a specified tolerance. This is achieved by estimating the size of the  $n+1$\textsuperscript{th} term using a least squares fit of the coefficients.

Extending the Taylor expansion to $m$ variables, an $n$\textsuperscript{th} order polynomial can be described by

\begin{equation}\label{eqn:ADS_vector_polynomial}
    \mathcal{T}_f(\bm{x}) = \sum_{|\beta| \leq n} a_\beta \bm{x}^{\beta}
\end{equation}
where multi-index notation is used for $\beta = \{\beta_1,...,\beta_{d}\}$ to denote the orders of the $m$ variables the trajectory is expanded with respect to and $\bm{x} = \{x_1,...,x_{d}\}$ is the vector of these variables, such as the initial position, initial velocity, etc. The size $S_i$ of the terms of order $i$ can then be found using an appropriate $L^p$-norm $||a_i||_p$ over all coefficients of $ |\beta| = i$. The most widely used are the $L^1$-norm (summation of absolute coefficients), $L^2$-norm (Euclidean), and $L^\infty$-norm (maximum of absolute coefficients). From the monotonicity of vector norms, it holds that $||a_i||_\infty \leq ||a_i||_2 \leq ||a_i||_1$. Therefore, for the most conservative estimate of coefficients sizes, the $L^1$-norm is selected, as this provides an upper bound.

\begin{equation}
    S_i = \sum_{\left|\beta\right| = i} \left| a_{\beta} \right|
\end{equation}

A least-squares fit of the exponential function

\begin{equation}\label{eqn:ADS_exponential_fit}
    E(i) = A \exp (B \cdot i)
\end{equation}
is then performed such that the determined coefficients $A$ and $B$ satisfy $E(i) = S_i$ for all non-zero values of $S_i$. Once obtained, the approximation error of the polynomial is easily estimated as the size of $E(n+1)=S_{n+1}$. By setting a tolerance for the maximum approximation error $e_\textrm{tol}$ one wishes to allow, a split will occur when the error exceeds this value. The splitting is therefore performed only when required and does not need to be known a priori. As the polynomials are multivariate, there remains the question of which variable to split along. This problem is automatically handled by the \gls{ADS} routine, which factors the coefficients with respect to each variable of the polynomial, and calculates corresponding absolute coefficient sizes and least-squares exponential fits. The sizes of the $n+1$\textsuperscript{th} terms are calculated, and the splitting direction is chosen according to the variable which contributes the largest approximation error. As a consequence, each split is guaranteed to have maximum impact. In the context of this work, the truncation error of the state on the event manifold $\bm{x}_{\textrm{e}}^*$ from \cref{eqn:Event_State_and_Time_Mapped} is checked with respect to the initial states $\bm{x}_0$, with splitting performed should the error exceed the set tolerance $e_{\textrm{tol}}$.

%###############################################################################
%###############################################################################

\section{Taylor Polynomial Bounders}\label{s:PolynomialBounding}
Consider an event map $\bm{\Phi} : \mathbb{R}^{d} \times \mathbb{R} \rightarrow \mathbb{R}^{d}$ that is approximated by Taylor polynomials. If the expansion point is given by $\overline{\bm{x}} \in \mathbb{R}^{d}$, such an approximation may be written as

\begin{equation}\label{eqn:Flow_Taylor_Approximation}
\bm{\Phi} (\bm{x},t) = \bm{P}_n(\bm{x} - \overline{\bm{x}}) + \bm{R}_n(\bm{x} - \overline{\bm{x}})
\end{equation}
where $\bm{P}_n : \mathbb{R}^{d} \rightarrow \mathbb{R}^{d}$ is the $n^{\textrm{th}}$ order polynomial and $\bm{R}_n: \mathbb{R}^{d} \rightarrow \mathbb{R}^{d}$ is the remainder term. These polynomials provide a continuous description of the system evolution, in contrast to the discrete samples obtained from Monte Carlo simulations. As such, all information regarding the event is encoded within the polynomials. This is a particularly powerful tool; however, the challenge lies in extracting this information without resorting to discrete sampling. A solution to this problem is polynomial bounding via interval arithmetic.

\subsection{Interval Arithmetic}
Instead of representing an input state as a single number, interval arithmetic represents the range of input states as an interval. Formally, this is defined as

\begin{equation}
    [a,b] = \{ x \in \mathbb{R} \, \mid \, a \leq x \leq b \}
\end{equation}
where $a$ and $b$ are the lower and upper bounds, respectively. These intervals are endowed with an algebra such that, for any number within the interval, performing the corresponding real arithmetic operation on those numbers results in a value that always lies within the interval obtained by applying the same operation to the intervals themselves. For two intervals $I_1 = [a_1,b_1]$ and $I_2 = [a_2,b_2]$, the elementary operators are defined by:

\begin{equation}\label{eqn:elementary_operators_of_intervals}
    \begin{aligned}
        I_1 + I_2 &= [a_1 + a_2, b_1 + b_2] \\
        I_1 - I_2 &= [a_1 - b_2, b_1 - a_2] \\
        I_1 \cdot I_2 &= [ \min\{a_1 a_2, a_1 b_2, b_1 a_2, b_1 b_2\} \, , \\
        & \quad \ \ \max\{a_1 a_2, a_1 b_2, b_1 a_2, b_1 b_2\} ]
        % \textrm{If} \, 0 \notin I_1, \: \frac{1}{I_1} &= \left[ \frac{1}{b_1},\frac{1}{a_1} \right]
    \end{aligned}
\end{equation}

The division operator is more complex, as it takes different forms depending on whether zero is within, at the edges, or outside the interval. Of course, many operations exist beyond the elementary ones, and for the complete arithmetic all operations for algebraic operations must be defined. However, for the purpose of analysing polynomial expressions, the algebra of \cref{eqn:elementary_operators_of_intervals} is sufficient.

Intervals can also be given in vector form, which shall be used extensively in the following sections of this manuscript. Formally,

\begin{equation}
    \begin{aligned}
    \bm{I} & = [\bm{a},\bm{b}] = \left[ (a_1, a_2, \dots, a_m), (b_1, b_2, \dots, b_m) \right] \\
    & = \left\{ \bm{x} \in \mathbb{R}^{d} \mid a_i \leq x_i \leq b_i\,, \; i=1,2,\cdots,m \,\right\}
    \end{aligned}
\end{equation}

Therefore, a separate interval exists for each element of the vector. In this way, $\bm{I}$ represents a vector of intervals, rather than an interval of vectors, a distinction that is crucial.

It should be noted that interval analysis suffers from two major drawbacks, known as the \textit{dependency problem} and \textit{wrapping effect}. The dependency problem occurs when the same interval appears multiple times within an expression. This may be illustrated with a simple example, whereby an arbitrary interval $I = [a,b]$ is subtracted from itself. Using the definition from \cref{eqn:elementary_operators_of_intervals}, one obtains $ I - I = [a-b,b-a]$. Clearly the true result for any quantity $x \in [a,b]$ should equal zero. Instead, each occurrence of the interval has been treated as independent, which leads to an overestimation of the true interval. Whilst some methods \cite{Dependency_Problem_Methods} exist to reduce overestimation automatically by approximating the dependency of intermediate results on the inputs, such methods add computational complexity and may not eliminate the inherent overestimation. The wrapping effect occurs when an interval greatly overestimates the true range of a function. For a simple illustration, consider the linear system $y=p$ and $x=p$ on the interval $p \in [-1,1]$. Using interval arithmetic, the bounds of both $x$ and $y$ are also $[-1,1]$. This results in a unit square $[-1,1]\times[-1,1]$, which suggests the system can take any value within this square, even though it is actually restricted to values on the line connecting $(-1,-1)$ and $(1,1)$. This effect fuels the long-term growth of integration errors. Various methods have been developed in the literature to address this issue, such as preconditioning \cite{BerzMakino_SupressionWrappingEffect_Preconditioning} and shrink wrapping \cite{BerzMakino_SupressionWrappingEffect_ShrinkWrapping}.

\subsection{Polynomial Bounding}
The Taylor polynomial $\bm{P}_n$, as given in \cref{eqn:Flow_Taylor_Approximation}, is a local approximation of the flow $\bm{\Phi}$ around the expansion point $\overline{\bm{x}}$. As the expansion point is fixed, interval arithmetic is not required prior to the Taylor approximation of the flow (i.e., $\overline{\bm{x}}$ need not be expressed as an interval, which would otherwise provide the potential interval of Taylor expansions over an interval of expansion points). Therefore, interval arithmetic may be applied directly to $\bm{P}_n$, which due to the polynomial form requires only the elementary operators defined in \cref{eqn:elementary_operators_of_intervals}. Interval analysis of the polynomial is conducted by considering individual monomials, which are then summed to obtain the full result. Without loss of generality, the system is scaled such that, for each $i \in \{1,\cdots,{d}\}$, $x_i -\overline{x_{i}} \in [-1,1]$ where $x_i$ and $\overline{x_{i}}$ are the $i^{\textrm{th}}$ components of $\bm{x}$ and $\overline{\bm{x}}$, respectively. This significantly simplifies the monomial calculation. Consider an arbitrary monomial $M_{\beta}$ with a defined multi-index $\beta = \{\beta_1,\cdots,\beta_{d}\}$ where $|\beta| \leq n$. The multi-index notation of $\bm{x}^\beta = x_1^{\beta_1} \cdots x_{d}^{\beta_{d}}$ is used for the exponents, where $\beta_i$ corresponds to the exponents of the variable $x_i$ in the monomials, $|\beta| = \beta_1 + \cdots + \beta_{d}$, and $\beta! = \beta_1! \cdots \beta_{d}!$. The monomial may be written as

\begin{equation}\label{eqn:Individual_Monomial}
    M_{\beta}(\bm{x} - \overline{\bm{x}}) = \frac{D^\beta \bm{f}(\overline{\bm{x}})}{\beta!} \prod_{i=1}^{d} (x_i - \overline{x_i})^{\beta_i}
\end{equation}

As the coefficients of the monomials do not depend upon $x_i$, they can be brought outside the product operator. As each $x_i - \overline{x_i} \in [-1,1]$, the product of any number of these intervals is also $[-1,1]$ by the rules of interval arithmetic defined in \cref{eqn:elementary_operators_of_intervals}. Therefore, one may easily bound any arbitrary monomial as

\begin{equation}\label{eqn:Individual_Monomial_Bounded}
    M_{\beta} \in \left[ -\left|\frac{D^\beta \bm{f}(\overline{\bm{x}})}{\beta!}\right|, \left|\frac{D^\beta \bm{f}(\overline{\bm{x}})}{\beta!}\right| \right]
\end{equation}

Accordingly, the resulting interval $\bm{I}_{\bm{P}}$ to bound the entire polynomial may be derived as

\begin{equation}\label{eqn:Polynomial_Interval}
    \begin{gathered}
        \forall \: \bm{x} - \overline{\bm{x}} \in [-\mathbf{1},\mathbf{1}] \, , \\
        \bm{P}_n(\bm{x}-\overline{\bm{x}}) \in \bm{I}_{\bm{P}} = \sum_{|\beta| \leq n} \left[ -\left|\frac{D^\beta \bm{f}(\overline{\bm{x}})}{\beta!}\right|, \left|\frac{D^\beta \bm{f}(\overline{\bm{x}})}{\beta!}\right| \right]
    \end{gathered}
\end{equation}

It is important to note that, while individual monomials do not suffer from the aforementioned dependency problem, the summation of monomials will. Nonetheless, the resulting interval $\bm{I}_{\bm{P}}$ will always enclose the polynomial, albeit with potential overestimation.

\begin{figure*}[!t]
    \centering
    \includegraphics[width=\textwidth]{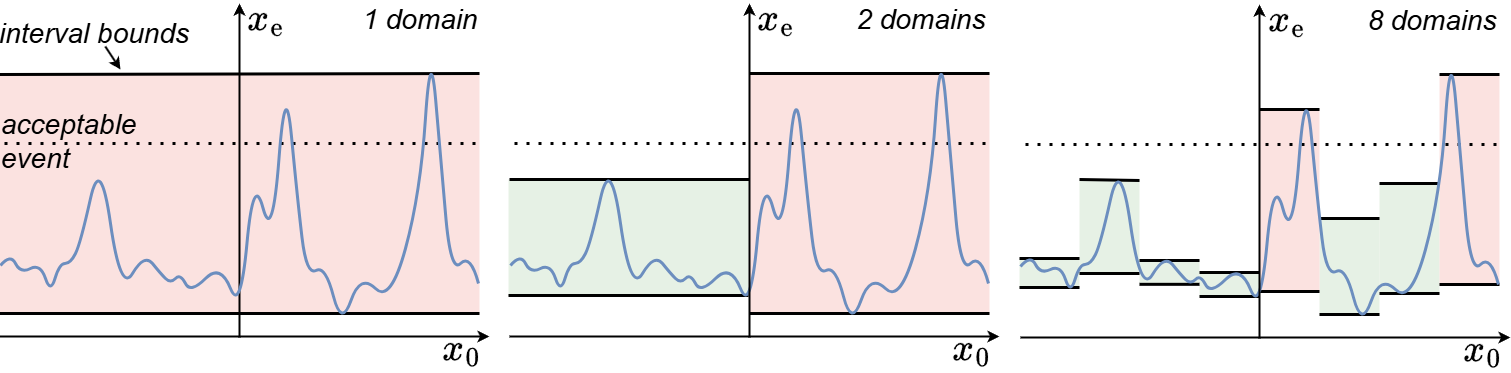}
	\caption{Illustration of how \gls{ADS} remedies the wrapping effect for a scalar event map $x_{\textrm{e}}(x_0)$. Solid horizontal lines represent the interval bounds, for increasing number of subdomains from left to right. Red and green shading represents intervals that are deemed unsafe and safe, respectively.}
	\label{fig:ADS_wrapping_problem}
\end{figure*}

As the Taylor approximation is a local one, the polynomials diverge quickly outside of the radius of convergence. Thus, the polynomials must be accurate over the given input interval on $\bm{x}$ for the bounding to be true. For this reason, in addition to the ability to analyse large uncertainty domains, \gls{ADS} is crucial. Without such a technique, the acquired bound on the event map would not be representative of the true events that are achieved, due to the polynomial divergence. Furthermore, \gls{ADS} serves a secondary purpose---reducing the wrapping effect. By splitting the domain into smaller subdomains, the enclosures around the state--space are localised, allowing for tighter bounding than applying a single bound to the entire domain. This is illustrated in \cref{fig:ADS_wrapping_problem} for an example scalar event map $x_0 \mapsto x_{\textrm{e}}$. Although the total range of the event across the entire state--space remains the same, a higher resolution of analysis is acquired as varying event intervals are obtained for different subdomains of the state--space. As depicted in \cref{fig:ADS_wrapping_problem}, 1 domain would provide a 0\% safety rating across the state--space, whereas 2 domains would provide 50\%, and 8 domains would provide 75\%. 
Thus, the safety of individual intervals within the state--space can be assessed, with others deemed unsafe, providing valuable insights.

\subsection{Remainder Bounding}
The methodologies employed within \gls{ADS} are now leveraged to bound the remainder in the Taylor approximation of the event map, $\bm{R}_{n}$. For each of the ${d}$ polynomials in $\bm{\Phi}$, the steps of \cref{eqn:ADS_vector_polynomial}-\cref{eqn:ADS_exponential_fit} are performed: a least-squares fit of the exponential function $E(i) = A \exp (B \cdot i)$ for all terms of order $i \leq n$ is performed such that the determined coefficients $A$ and $B$ satisfy $E(i) = S_i$ for all non-zero values of polynomial coefficients $S_i$. For the most conservative estimate, the $L^1$-norm is again selected to provide an upper bound of the coefficient sizes at each order. If the domain is split sufficiently, each subdomain will be within the convergence radius of the corresponding Taylor approximations, and the coefficients of the polynomials will in fact decay exponentially. This is a direct consequence of Taylor’s
Theorem~\cite{ADS_original}. As the use of the $L^1$-norm provides an upper bound of the magnitude of the terms of each order, the corresponding extrapolation to the $n+1^\textrm{th}$ order also provides an upper bounding.

Finally, the vector $\tilde{\bm{R}}_{n} \in \mathbb{R}_{+}^{d}$ is assembled, containing the extrapolated coefficients $E(n+1)$ for each of the ${d}$ polynomials of the map, and therefore the estimated remainder error. Thus, the bounds of the event map are given as

\begin{equation}
    \bm{\Phi}(\bm{x},t) \in \bm{I}_{\bm{P}} + \left[-\tilde{\bm{R}}_{n},\tilde{\bm{R}}_{n}\right]
\end{equation}

The methodology outlined in this section provides a framework for bounding any event map approximated via Taylor polynomials. This approach is applied to two specific scenarios in the subsequent sections.

%###############################################################################
%###############################################################################

\section{Relative Orbital Motion}\label{s:ClohessyWiltshire}

\begin{figure*}[!t]
\centering
    \begin{subfigure}{0.5\textwidth}
    \centering
    \def\layersep{2.5cm}
    \def\hiddennum{14}
    \def\inputnum{4}
    \def\outputnum{2}
    \def\verticalscale{0.6}

    \resizebox{\textwidth}{!}{%
    \begin{tikzpicture}[shorten >=1pt,->,draw=black!50, node distance=\layersep]
        \tikzstyle{every pin edge}=[<-,shorten <=1pt]
        \tikzstyle{neuron}=[circle,fill=black!25,minimum size=10pt,inner sep=0pt]
        \tikzstyle{input neuron}=[neuron, fill=gray!50];
        \tikzstyle{output neuron}=[neuron, fill=green!50];
        \tikzstyle{hidden neuron}=[neuron, fill=blue!50];
        \tikzstyle{annot} = [text width=4em, text centered]
    
        % Draw the input layer nodes
        \node[input neuron, pin=left:\LARGE$x$]   (Input-1) at (0,-2*\verticalscale) {};
        \node[input neuron, pin=left:\LARGE$y$]   (Input-2) at (0,-3*\verticalscale) {};
        \node[input neuron, pin=left:\LARGE$v_x$] (Input-3) at (0,-4*\verticalscale) {};
        \node[input neuron, pin=left:\LARGE$v_y$] (Input-4) at (0,-5*\verticalscale) {};
    
        % Draw the hidden layer nodes
        \foreach \i in {1,...,\hiddennum}
            \path[yshift=(\hiddennum-\inputnum)*2.95*\verticalscale mm]
                node[hidden neuron] (Hidden1-\i) at (\layersep,-\i*0.8*\verticalscale) {};
    
        \foreach \i in {1,...,\hiddennum}
            \path[yshift=(\hiddennum-\inputnum)*2.95*\verticalscale mm]
                node[hidden neuron] (Hidden2-\i) at (2*\layersep,-\i*0.8*\verticalscale) {};

        \foreach \i in {1,...,\hiddennum}
            \path[yshift=(\hiddennum-\inputnum)*2.95*\verticalscale mm]
                node[hidden neuron] (Hidden3-\i) at (3*\layersep,-\i*0.8*\verticalscale) {};
   
        % Draw the output layer node
        \node[output neuron,pin={[pin edge={->}]right:\LARGE$\alpha_x$}, right of=Hidden3-1] (Output-1) at (3*\layersep,-3*\verticalscale) {};
        \node[output neuron,pin={[pin edge={->}]right:\LARGE$\alpha_y$}, right of=Hidden3-2] (Output-2) at (3*\layersep,-4*\verticalscale) {};

        \node at ($(Hidden2-8)!.5!(Hidden3-8)$) {\huge\ldots};
    
        % Connect every node in the input layer with every node in the hidden1 layer.
        \foreach \source in {1,...,\inputnum}
            \foreach \dest in {1,...,\hiddennum}
                \path (Input-\source) edge (Hidden1-\dest);
    
        \foreach \source in {1,...,\hiddennum}
            \foreach \dest in {1,...,\hiddennum}
                \path (Hidden1-\source) edge (Hidden2-\dest);
    
        % Connect every node in the hidden2 layer with the output layer
        \foreach \source in {1,...,\hiddennum}{
        	\foreach \dest in {1,...,\outputnum}{
        		\path (Hidden3-\source) edge (Output-\dest);
        	}
        }

        % Annotate the layers
        \node[annot,above of=Hidden1-1, node distance=0.55cm] (h1) {\Large SIREN};
        \node[annot,above of=Hidden2-1, node distance=0.55cm] (h2) {\Large SIREN};
        \node[annot,above of=Hidden3-1, node distance=0.55cm] (h3) {\Large SIREN};
        \node[annot,right of=h3] {\Large linear};
    \end{tikzpicture}%
    }
    \begin{minipage}{.1cm}
            \vfill
    \end{minipage}
    \end{subfigure}%
    \begin{subfigure}{0.5\textwidth}
        \centering
        \includegraphics[width=0.9\textwidth]{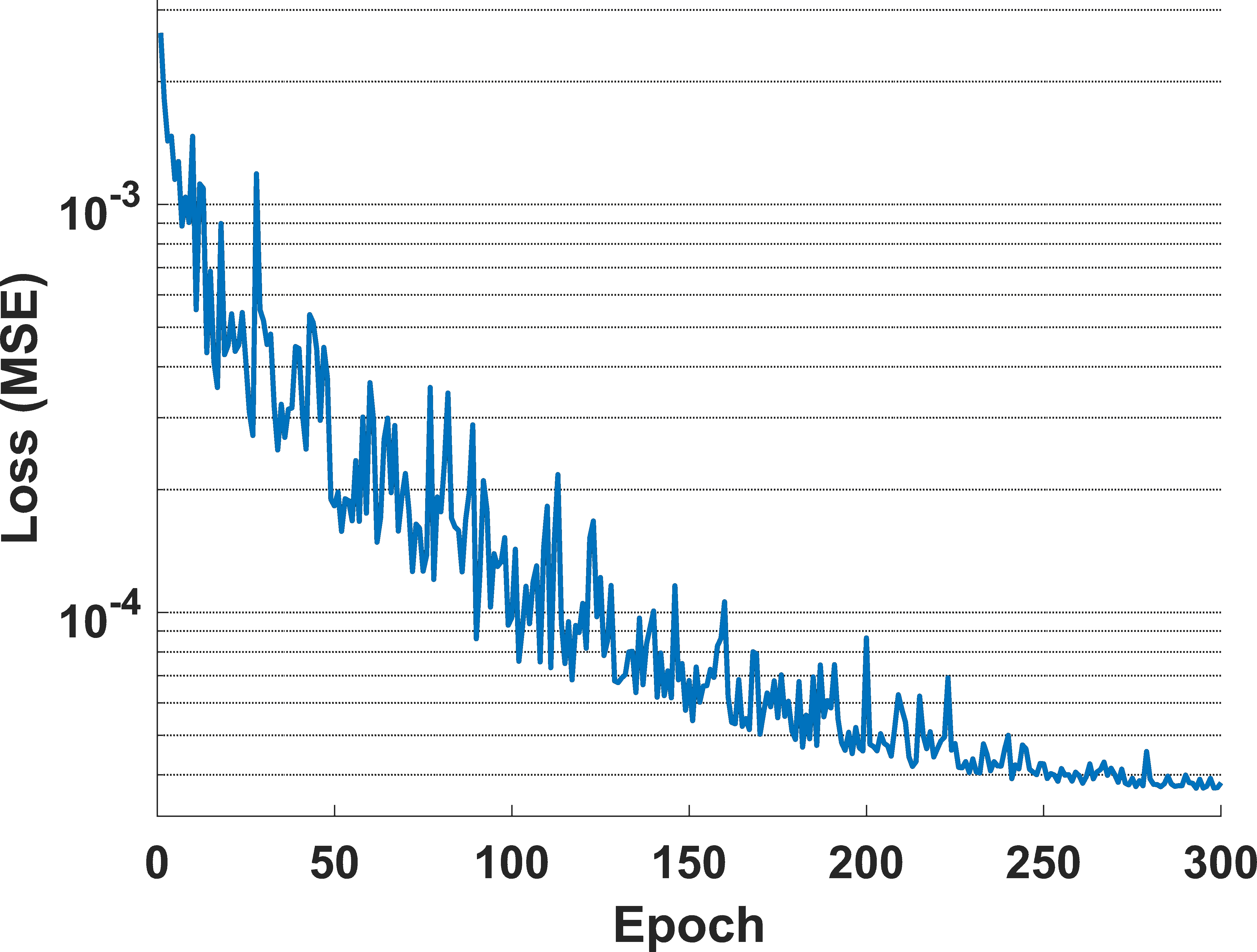}
    \end{subfigure}
\caption{Neural network to replicate the optimal control direction for the Clohessy--Wiltshire scenario, (left) network architecture and (right) losses during training.}
\label{fig:CW_NN}
\end{figure*}
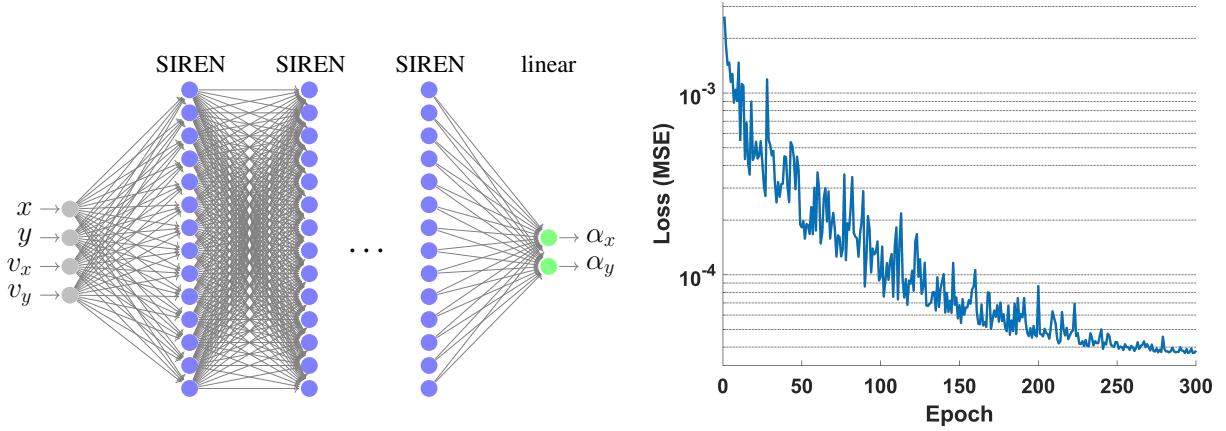  

\subsection{Deep Neural Network}\label{ss:ClohessyWiltshire_DNN}
Ref.~\cite{MyTimeOptPaper} investigated a time-optimal control problem for relative orbital motion using the Clohessy--Wiltshire equations, with the objective of achieving rendezvous with a target craft located at the centre of the reference frame. While that study utilized Taylor methods similar to those described here to build polynomial guidance maps, the present work revisits this scenario by training a \gls{DNN} to learn the mapping $\bm{x} \mapsto \bm{\alpha}^*$ where $\bm{x} \in \mathbb{R}^4$ is the state and $\bm{\alpha}^* \in \mathbb{R}^2$ is the optimal control direction. In fact, the same polynomial map derived in the aforementioned work is utilised to rapidly generate time-optimal trajectories for the database, in a similar manner to that in Ref. \cite{MyDNNPaper}. Each trajectory was split into $k$ equal segments over its corresponding time of flight. A continuous uniform distribution was then used to sample a random state from within each segment. Three datasets were created: a training set $\mathcal{D}_{\textrm{train}}$ with $k=100$ consisting of 10,000,000 samples; a validation set $\mathcal{D}_{\textrm{val}}$ with $k=100$ consisting of 100,000 samples; and a test set $\mathcal{D}_{\textrm{test}}$ with $k=1$ consisting of 1000 samples.

As previously mentioned, the \gls{DNN} must be a smooth function to allow expansion via Taylor polynomials. Recently, \glspl{SIREN} have emerged as a powerful choice of activation function. Studies have demonstrated that \glspl{SIREN} are both effective and versatile in representing images and videos, as well as solving \glspl{TPBVP} \cite{SIREN_Original}. Moreover, they have been successfully applied to replicate optimal guidance policies in spaceflight applications with promising results \cite{SIREN_G&CNET}. Not only may \gls{SIREN} layers outperform other typical activation functions in certain scenarios, but more importantly, such layers are completely smooth, making them an ideal choice for the proposed network. 

For the $i^{\textrm{th}}$ layer, denoted as $\bm{y}_i : \mathbb{R}^{M_i} \rightarrow \mathbb{R}^{N_i}$, consisting of $N_i$ neurons and taking an input vector $\bm{x}_i \in \mathbb{R}^{M_i}$, the \gls{SIREN} layer is represented as

\begin{equation}
\bm{y}_i = \sin\left(\omega \left(\mathbf{W}_i \bm{x}_i + \bm{b}_i\right) \right)
\end{equation}
where $\mathbf{W}_i \in \mathbb{R}^{N_i \times M_i}$ and $\bm{b}_i \in \mathbb{R}^{N_i}$ are the weight matrix and bias vector, respectively. The frequency parameter $\omega$ may either be a learnable parameter or fixed, allowing the model to capture higher or lower frequency features in the dataset for larger or smaller values, respectively.

It is important to note that the objective of this work is not to train the best \gls{DNN} possible for the given task, but rather to develop and demonstrate a new methodology of obtaining safety assessments for feedback controllers. As such, a comprehensive investigation of network architectures was not conducted; however, a brief analysis was performed to develop a controller that achieved satisfactory performance. The \gls{DNN} architecture consisted of an input layer with 4 neurons, 5 hidden layers each with 32 neurons, and an output layer with 2 neurons. \glspl{SIREN} were used for all activation layers. A specific weight initialisation scheme was applied, as detailed in Ref. \cite{SIREN_Original}. For the first layer, the weights were uniformly distributed such that $\mathbf{W}_1 \sim \mathcal{U}(-1/M_1,1/M_1)$. The remaining layers were initialised such that $\mathbf{W}_i \sim \mathcal{U}(-\sqrt{6/M_i}/\omega,\sqrt{6/M_i}/\omega)$. Additionally, the data was scaled such that the inputs $\bm{x}_1 \sim \mathcal{U}(-1,1)$. The number of epochs was set to 300, and the batch size to 1000. The initial learning rate was set to $1\times10^{-4}$, decreasing by a factor of 0.6 if the loss calculated on $\mathcal{D}_{\textrm{val}}$ did not improve by 1$\%$ over 10 epochs. The loss function was the mean squared error between the predicted $\bm{\alpha}$ and corresponding optimal values. Finally, AdamW without AMSgrad was chosen as the optimiser. The neural network architecture and the losses calculated during training, evaluated on $\mathcal{D}_{\textrm{val}}$, are displayed in \cref{fig:CW_NN}.

%###############################################################################

\subsection{Safety Analysis}

\begin{figure*}[!t]
    \centering
    \includegraphics[width=\textwidth]{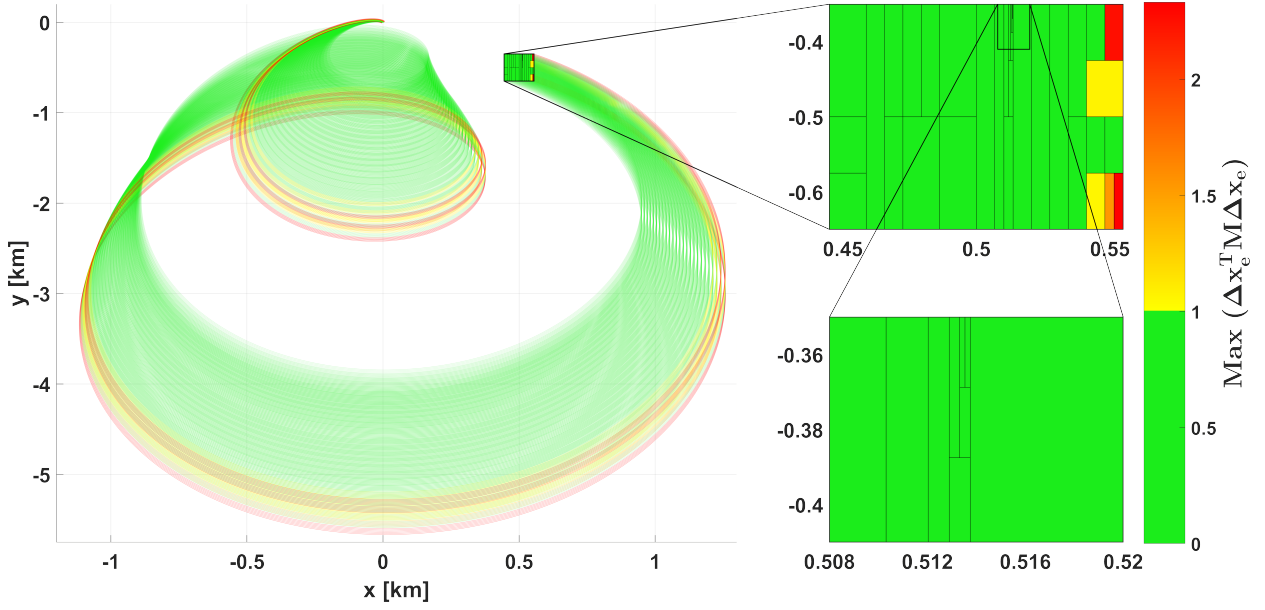}
	\caption{Plot of the event map bounding for the Clohessy--Wiltshire scenario using the trained \gls{DNN} controller. A colour bar is applied to indicate safe and unsafe subdomains, based on the bounds of the squared length.} 
	\label{fig:CW_NN_safety_map}
\end{figure*}

The trained \gls{DNN} was then embedded into the dynamical \glspl{ODE}, making it ready for analysis. For this scenario, the event of interest was the minimum squared length $L = \Delta\bm{x}^\top \cdot \mathbf{M} \cdot \Delta\bm{x} \in \mathbb{R}_{+}$, which represents the closest approach within the state-space along a given trajectory. Here, $\Delta\bm{x} \in \mathbb{R}^4$ is the residual between the spacecraft state and target state, and $\mathbf{M} \in \mathbb{R}^{4\times4}_{+}$ is a positive-definite weighting matrix used to normalise the position and velocity components of the state residual. Since the target for rendezvous was positioned at the centre of the reference frame with zero velocity, the state and the state residual are equivalent. 

The system was propagated and at each integration step $L$ was calculated. If the value was the lowest obtained so far along the current trajectory, it was stored in memory along with the time at the current step, denoted by $\overline{t}_{\textrm{e}}$. This continued until the pre-set maximum time of flight of 4 hours was reached (for reference, the nominal time of flight was previously found to be 3.34 hours). Thus, the minimum $L$ was detected over the course of the trajectory, albeit with low accuracy due to overstepping. $\overline{t}_{\textrm{e}}$ was then refined and subsequently mapped using the algorithms outlined in \cref{ss:EventMap} to obtain the event map $\bm{x}_{\textrm{e}}^*$. The \gls{ADS} algorithm then analysed the truncation error on these polynomials, and performed the necessary splits. The parameters for \gls{ADS} included: an initial domain of interest in position only, defined by $\delta \bm{r}_0 \in \pm [55,150]^\top$ m; a polynomial order of 4; a maximum of $N_{\textrm{max}} = 15$ splits to create a single sub-domain; and a splitting tolerance of $e_{\textrm{tol}} = 1\times10^{-4}$.

The resulting event map consisted of 35 subdomains, which were subsequently analysed via polynomial bounding. For each subdomain, the squared length $L = \Delta{\bm{x}_{\textrm{e}}^*}^\top \cdot \, \mathbf{M} \, \cdot \, \Delta\bm{x}_{\textrm{e}}^*$ was calculated. For an acceptable squared length, a position residual of 5 m and velocity residual of 0.01 m/s was set, yielding the following metric:

\begin{equation}
    \mathbf{M} = \frac{1}{2}
    \begin{bmatrix}
        \dfrac{1}{5^2} & 0 & 0 & 0 \\
        0 & \dfrac{1}{5^2} & 0 & 0 \\
        0 & 0 & \dfrac{1}{0.01^2} & 0 \\
        0 & 0 & 0 & \dfrac{1}{0.01^2} \\
    \end{bmatrix}
\end{equation}

An acceptable event then corresponded to an $L \leq 1$ across the entire subdomain. The results, presented in \cref{fig:CW_NN_safety_map}, illustrate the safety analysis for the developed \gls{DNN} controller. The colour bar indicates whether a subdomain is safe or unsafe, based on the results of the bounding. Of the 35 subdomains, 5 were deemed unacceptable, with the maximum squared length found to be 2.33 across the entire domain. It was observed that any uncertainty in the domain $x\in[0.45,0.54125]$ km, $y\in[-0.65,-0.35]$ km was assessed as safe, with maximum squared lengths below the acceptable threshold. The worst results occurred at the extremal values of the initial uncertainty domain, specifically for $x > 0.54125$ km in both the $\pm y$-directions. Overall, 92.2\% of the initial state--space was assessed as safe. It is worth noting that a higher resolution analysis can be achieved by reducing $e_{\textrm{tol}}$ and increasing $N_{\textrm{max}}$. These results demonstrate that the proposed method can effectively assess the safety of individual intervals when using the trained \gls{DNN} controller, offering valuable insights for operational decision-making.

%###############################################################################
%###############################################################################

\section{Earth--Mars Transfer}\label{s:Earth_Mars}
\subsection{Deep Neural Network}

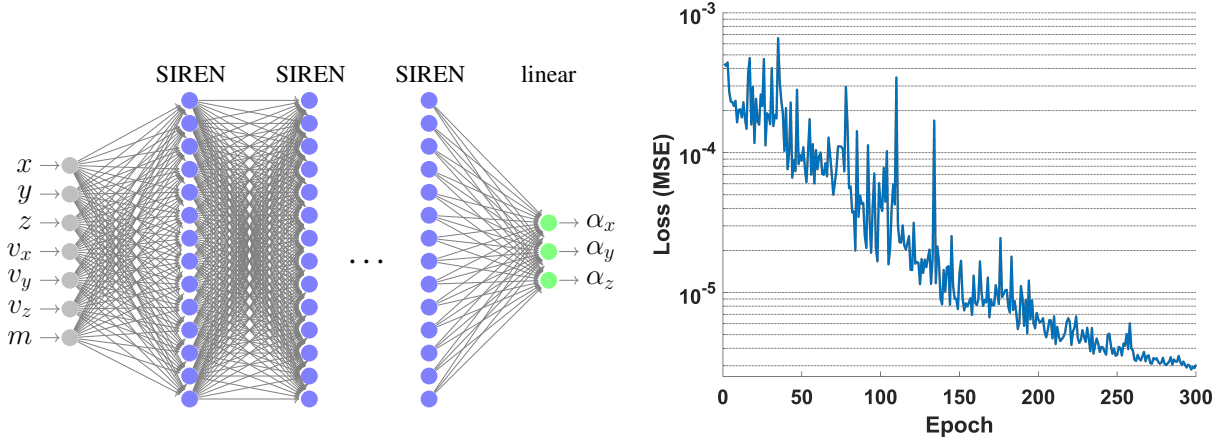
\begin{figure*}
\centering
    \begin{subfigure}{0.5\textwidth}
    \centering
    \def\layersep{2.5cm}
    \def\hiddennum{14}
    \def\inputnum{7}
    \def\outputnum{3}
    \def\verticalscale{0.6}

    \resizebox{\textwidth}{!}{%
    \begin{tikzpicture}[shorten >=1pt,->,draw=black!50, node distance=\layersep]
        \tikzstyle{every pin edge}=[<-,shorten <=1pt]
        \tikzstyle{neuron}=[circle,fill=black!25,minimum size=10pt,inner sep=0pt]
        \tikzstyle{input neuron}=[neuron, fill=gray!50];
        \tikzstyle{output neuron}=[neuron, fill=green!50];
        \tikzstyle{hidden neuron}=[neuron, fill=blue!50];
        \tikzstyle{annot} = [text width=4em, text centered]
    
        % Draw the input layer nodes
        \node[input neuron, pin=left:\LARGE$x$]     (Input-1) at (0,-1*\verticalscale) {};
        \node[input neuron, pin=left:\LARGE$y$]     (Input-2) at (0,-2*\verticalscale) {};
        \node[input neuron, pin=left:\LARGE$z$]     (Input-3) at (0,-3*\verticalscale) {};
        \node[input neuron, pin=left:\LARGE$v_x$]   (Input-4) at (0,-4*\verticalscale) {};
        \node[input neuron, pin=left:\LARGE$v_y$]   (Input-5) at (0,-5*\verticalscale) {};
        \node[input neuron, pin=left:\LARGE$v_z$]   (Input-6) at (0,-6*\verticalscale) {};
        \node[input neuron, pin=left:\LARGE$m$]     (Input-7) at (0,-7*\verticalscale) {};
    
        % Draw the hidden layer nodes
        \foreach \i in {1,...,\hiddennum}
            \path[yshift=(\hiddennum-\inputnum)*2.95*\verticalscale mm]
                node[hidden neuron] (Hidden1-\i) at (\layersep,-\i*0.8*\verticalscale) {};
    
        \foreach \i in {1,...,\hiddennum}
            \path[yshift=(\hiddennum-\inputnum)*2.95*\verticalscale mm]
                node[hidden neuron] (Hidden2-\i) at (2*\layersep,-\i*0.8*\verticalscale) {};

        \foreach \i in {1,...,\hiddennum}
            \path[yshift=(\hiddennum-\inputnum)*2.95*\verticalscale mm]
                node[hidden neuron] (Hidden3-\i) at (3*\layersep,-\i*0.8*\verticalscale) {};
   
        % Draw the output layer node
        \node[output neuron,pin={[pin edge={->}]right:\LARGE$\alpha_x$}, right of=Hidden3-1] (Output-1) at (3*\layersep,-3*\verticalscale) {};
        \node[output neuron,pin={[pin edge={->}]right:\LARGE$\alpha_y$}, right of=Hidden3-2] (Output-2) at (3*\layersep,-4*\verticalscale) {};
        \node[output neuron,pin={[pin edge={->}]right:\LARGE$\alpha_z$}, right of=Hidden3-3] (Output-3) at (3*\layersep,-5*\verticalscale) {};

        \node at ($(Hidden2-8)!.5!(Hidden3-8)$) {\huge\ldots};
    
        % Connect every node in the input layer with every node in the hidden1 layer.
        \foreach \source in {1,...,\inputnum}
            \foreach \dest in {1,...,\hiddennum}
                \path (Input-\source) edge (Hidden1-\dest);
    
        \foreach \source in {1,...,\hiddennum}
            \foreach \dest in {1,...,\hiddennum}
                \path (Hidden1-\source) edge (Hidden2-\dest);
    
        % Connect every node in the hidden2 layer with the output layer
        \foreach \source in {1,...,\hiddennum}{
        	\foreach \dest in {1,...,\outputnum}{
        		\path (Hidden3-\source) edge (Output-\dest);
        	}
        }

        % Annotate the layers
        \node[annot,above of=Hidden1-1, node distance=0.55cm] (h1) {\Large SIREN};
        \node[annot,above of=Hidden2-1, node distance=0.55cm] (h2) {\Large SIREN};
        \node[annot,above of=Hidden3-1, node distance=0.55cm] (h3) {\Large SIREN};
        \node[annot,right of=h3] {\Large linear};
    \end{tikzpicture}%
    }
    \begin{minipage}{.1cm}
            \vfill
    \end{minipage}
    \end{subfigure}%
    \begin{subfigure}{0.5\textwidth}
         \centering
        \includegraphics[width=0.9\textwidth]{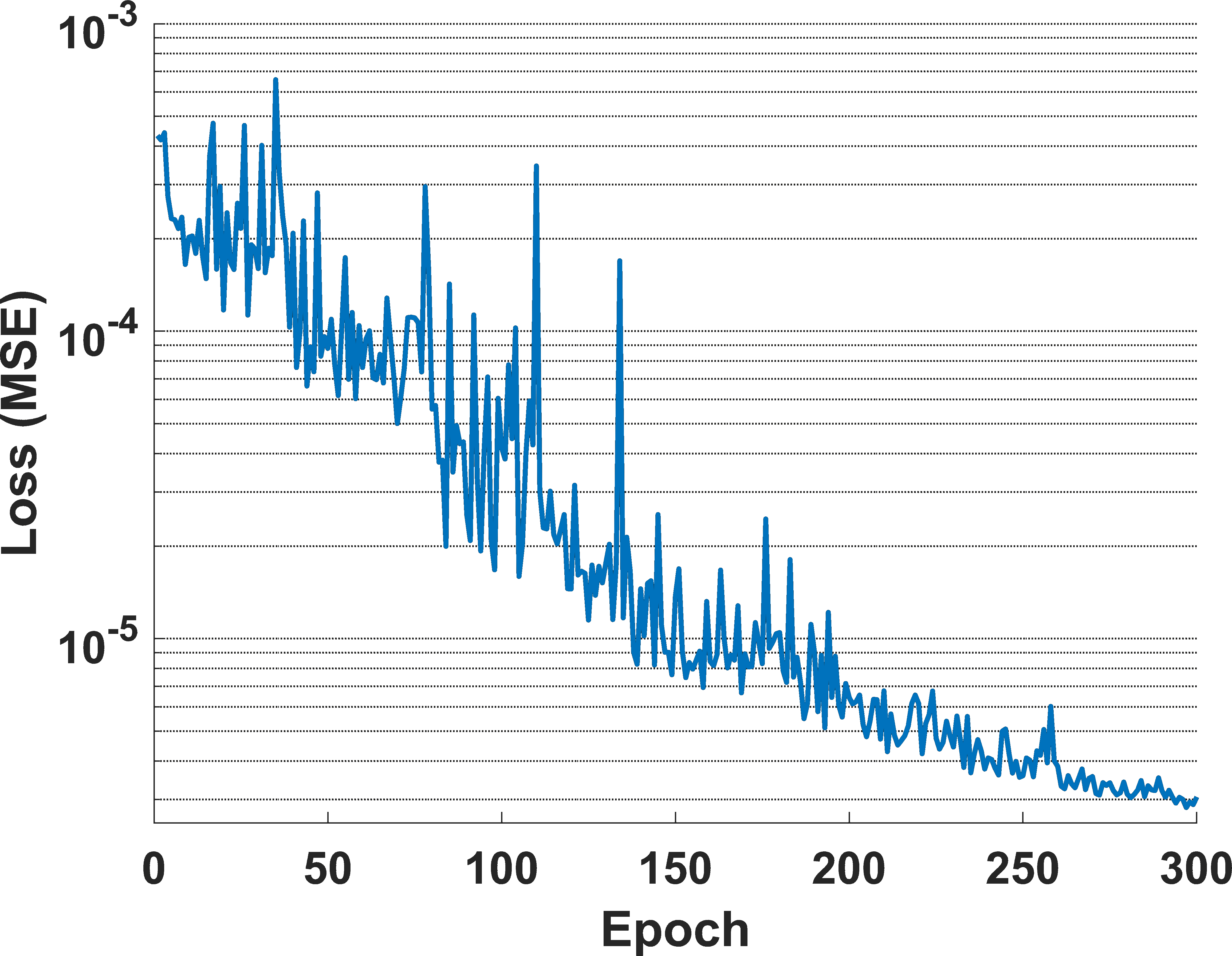}
    \end{subfigure}
\caption{Neural network to replicate the optimal control direction for the Earth--Mars transfer, (left) network architecture and (right) losses during training.}
\label{fig:EarthMars_NN}
\end{figure*}  

% \begin{figure}[!b]
%     \centering
%     \includegraphics[width=0.65\textwidth]{Figures/EarthMars_NN_losses.png}
%     \caption[Losses during the training of the neural network for the Earth--Mars transfer scenario to replicate the optimal control direction]{Losses during the training of the neural network for the Earth--Mars transfer scenario to replicate the optimal control direction.}
% 	\label{fig:EarthMars_NN}
% \end{figure}

The second case study focuses on the time-optimal Earth--Mars transfer studied in Ref.~\cite{MyTimeOptPaper}. In this problem, a spacecraft departs Earth at epoch 8510 MJD2000 to perform a rendezvous with Mars after a nominal time of flight of 1113.812 days. As with the previous example, a \gls{DNN} is trained to approximate the mapping $\bm{x} \mapsto \bm{\alpha}^*$, where $\bm{x} \in \mathbb{R}^7$ denotes the state and $\bm{\alpha}^* \in \mathbb{R}^3$ the optimal control direction. The same polynomial map derived in the aforementioned work is utilised to rapidly generate time-optimal trajectories for the database. Each trajectory was split into $k$ equal segments over its corresponding time of flight, with a uniform distribution used to sample a random state from within each segment. Three datasets were created: a training set $\mathcal{D}_{\textrm{train}}$ with $k=500$ consisting of 10,000,000 samples; a validation set $\mathcal{D}_{\textrm{val}}$ with $k=100$ consisting of 100,000 samples; and a test set $\mathcal{D}_{\textrm{test}}$ with $k=1$ consisting of 1000 samples. The \gls{DNN} architecture consisted of an input layer with 7 neurons, 5 hidden layers each with 64 neurons, and an output layer with 3 neurons. \glspl{SIREN} were used for all activation layers, and the same weight initialisation scheme as in \cref{ss:ClohessyWiltshire_DNN} was performed. The number of epochs was set to 300, and the batch size to 1000. The initial learning rate was set to $1\times10^{-4}$, decreasing by a factor of 0.6 if the loss calculated on $\mathcal{D}_{\textrm{val}}$ did not improve by 1$\%$ over 10 epochs. The loss function was the mean squared error between the predicted $\bm{\alpha}$ and corresponding optimal values. Finally, AdamW without AMSgrad was chosen as the optimiser. The neural network architecture and the losses calculated during training, evaluated on $\mathcal{D}_{\textrm{val}}$, are displayed in \cref{fig:EarthMars_NN}.

%###############################################################################

\subsection{Safety Analysis}
The trained \gls{DNN} was then embedded into the dynamical \glspl{ODE}. For this scenario, the event of interest was the relative velocity upon entering the \gls{SOI} of Mars, given by ${\left|\left|  \bm{v} - \bm{v}_{\textrm{M}} \right|\right| \big|_{\left|\left| \mathbf{r}\right|\right| = r_{\textrm{SOI}}} \in \mathbb{R}_{+}}$, where $r_{\textrm{SOI}} = 5.77\times 10^{5}$ km is the radius of the \gls{SOI}, with $\bm{v} \in \mathbb{R}^3$ and $\bm{v}_{\textrm{M}} \in \mathbb{R}^3$ denote the velocities of the spacecraft and Mars, respectively. 

The system was propagated and at each integration step the relative position $\left|\left|  \bm{r} - \bm{r}_{\textrm{M}} \right|\right| $ was computed, where $\bm{r} \in \mathbb{R}^3$ and $\bm{r}_{\textrm{M}} \in \mathbb{R}^3$ denote the positions of the spacecraft and Mars, respectively. If the condition $\left|\left|  \bm{r} - \bm{r}_{\textrm{M}} \right|\right| \leq r_{\textrm{SOI}}$ was met, the current time, denoted as $\overline{t}_{\textrm{e}}$, was stored in memory and propagation was terminated. Otherwise, propagation continued until the pre-set maximum time of flight of 1336.6 days was reached (for reference, the nominal time of flight was previously found to be 1113.8 days, hence the maximum time-of-flight allowed is set to 1.2 times the nominal). Thus, the time at which the spacecraft entered the \gls{SOI} was detected, albeit with low accuracy due to overstepping. This event time was then refined and mapped using the algorithms outlined in \cref{ss:EventMap}, yielding $t_{\textrm{e}}^*$. Since the mapping corresponded to a fixed norm of position---that of the \gls{SOI}---positional elements were omitted from the event map. Instead, the output consisted solely of the relative velocity at the \gls{SOI}, denoted as $\Delta \bm{v}_{\textrm{e}} \in \mathbb{R}^3$, and the associated time of flight, $t_{\textrm{e}}^*$. \gls{ADS} was then used to analyse the truncation error on these polynomials, for which the parameters included: an initial domain in position only, consisting of $\delta \bm{r}_0 \in \pm [1.25,1.25,0.25]^\top \times 10^5$ km; a polynomial order of 4; a maximum of $N_{\textrm{max}} = 12$ splits to create a single sub-domain; and a splitting tolerance of $e_{\textrm{tol}} = 1\times10^{-8}$.

\begin{figure*}
    \centering
    \includegraphics[width=0.95\textwidth]{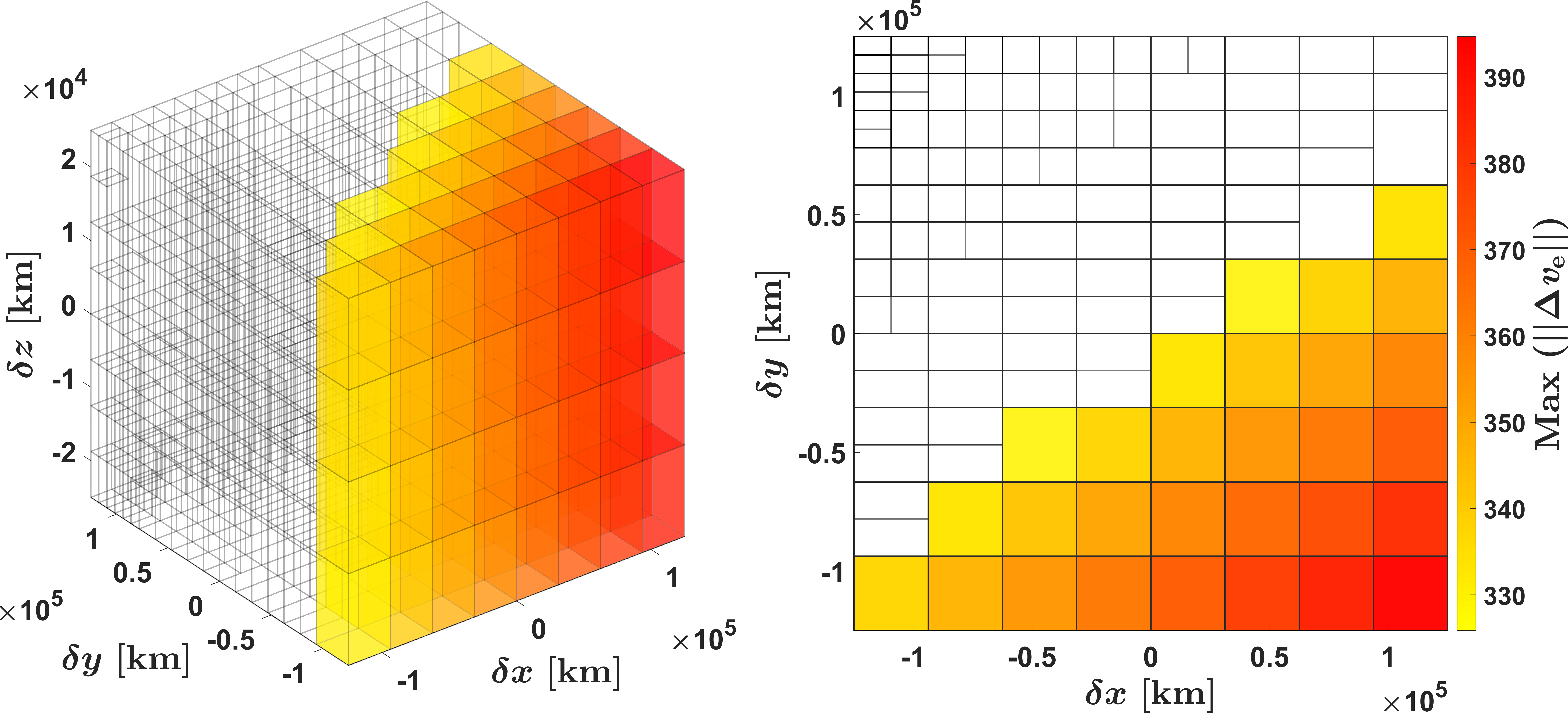}
	\caption{Plots of the event map bounding for the Earth--Mars scenario, including (left) 3D and (right) top-down views. Transparent subdomains are assessed as safe. A colour bar is applied to unsafe subdomains to indicate the severity of the violation.} 
	\label{fig:EarthMars_NN_safety_map}
\end{figure*}

The resulting event map consisted of 478 subdomains, which were subsequently analysed via polynomial bounding. For each subdomain, the norm of the relative velocity at the event $\left|\left| \Delta \bm{v}_{\textrm{e}} \right|\right|$ was calculated. To determine an acceptable threshold for the relative velocity, the polynomial map from~\cite{MyTimeOptPaper} was once again utilised. A Monte Carlo simulation was performed, selecting 10,000 samples from the initial domain of the map, which produced optimal trajectories. The aforementioned detection and refinement algorithm was then applied to accurately obtain the relative velocities at the \gls{SOI}, which consequently corresponded to optimal values. The maximum value across all samples was found to be 296.25 m/s, to which an additional tolerance of 10\% was added to define the threshold for the maximum accepted relative velocity. The event map was then analysed based on this criterion, with the results presented in \cref{fig:EarthMars_NN_safety_map}. 

The results indicated clear trends: the safety of subdomains was largely dictated by errors in the $x$-$y$ plane; errors experienced in the $-x$ and $+y$ directions remained within safe bounds; and the velocity tolerance was increasingly violated for errors in the $+x$ and $-y$ directions. Moreover, any infinitesimal perturbation from the nominal in the $+x$ and $-y$ directions moved the spacecraft into a subdomain in which the velocity tolerance at the \gls{SOI} exceeded the set threshold. Given these findings, one could argue that the controller is inadequate for safe operation, due to the large area of the state--space deemed unsafe, as well as its sensitivity to nominal conditions that could place the spacecraft in an unsafe region. Thus, for safe operations, such a controller may need to be adapted, re-trained, and re-assessed using the proposed methodology until an acceptable controller is obtained.

%###############################################################################
%###############################################################################

\section{Conclusion}
This work developed a methodology for assessing the safety of smooth feedback controllers via Taylor polynomial bounding across extensive state--space domains. By embedding the controller within the dynamical \glspl{ODE}, the system’s flow was approximated by Taylor polynomials, which were subsequently manipulated to compute quantities of interest associated with specific events. Such quantities, known as event maps, projected an initial state domain onto an event manifold. Polynomial bounding techniques were then employed to determine the possible range of event outcomes and assess their adherence to predefined safety tolerances. \gls{ADS} played a crucial role in ensuring the accuracy of the resulting polynomials while mitigating the wrapping effect, enabling a precise, subdomain-level analysis. The methodology was applied to trained regression \gls{DNN} controllers in two different scenarios, culminating in maps across the state domain that classified individual subdomains as either safe or unsafe based on the predefined tolerances. In this way, the resulting domain map exhibited similarities to a heat map, a widely used tool in the field of explainable AI. This approach provides a systematic framework for bounding the outcomes of guidance operations, enabling robust safety assessment. The results demonstrate the potential of the methodology to assess the safety of neural network–based controllers and to support confidence in their real-world deployment.

% \section*{ACKNOWLEDGMENT}
% The authors acknowledge members of the Advanced Concepts Team at the European Space Agency for the helpful discussions regarding this research.

\bibliographystyle{IEEEtran}
\bibliography{references}

\end{document}